\documentclass[10pt,conference]{IEEEtran}
\usepackage[utf8]{inputenc}
\usepackage{multirow}
\usepackage{tabularx}
\usepackage{booktabs} 
\usepackage{hyperref}
\usepackage{amsmath}
\usepackage{verbatim}
\usepackage{siunitx}
\usepackage[shortlabels]{enumitem}
\usepackage{amsmath}
\usepackage{fancybox}
\usepackage{url}
\usepackage{graphicx}
\usepackage{setspace}
\usepackage{soul}
\usepackage{multirow}
\usepackage[roman]{parnotes}
\usepackage{color}
\usepackage{colortbl}
\usepackage{listings}

\newcounter{observation}
\newcommand{\observation}[1]{\refstepcounter{observation}
	\begin{center}
		\Ovalbox{
			\begin{minipage}{0.93\columnwidth}
				{\bf Observation \arabic{observation}:} #1
			\end{minipage}
		}
	\end{center}
}
\newcolumntype{L}{>{\centering\arraybackslash}m{3cm}}

\title{We'll Fix It in Post: What Do Bug Fixes in Video Game Update Notes Tell Us?}

\begin{document}

\author{\IEEEauthorblockN{Andrew Truelove}
\IEEEauthorblockA{\textit{University of California, Irvine}\\
Irvine, CA, USA \\
truelova@uci.edu}
\and
\IEEEauthorblockN{Eduardo Santana de Almeida}
\IEEEauthorblockA{\textit{Federal University of Bahia}\\
Salvador, Brazil \\
esa@rise.com.br}
\and
\IEEEauthorblockN{Iftekhar Ahmed}
\IEEEauthorblockA{\textit{University of California, Irvine}\\
Irvine, CA, USA \\
iftekha@uci.edu}

}

\maketitle
\begin{abstract}
Bugs that persist into releases of video games can have negative impacts on both developers and users, but particular aspects of testing in game development can lead to difficulties in effectively catching these missed bugs. It has become common practice for developers to apply updates to games in order to fix missed bugs. These updates are often accompanied by notes that describe the changes to the game included in the update. However, some bugs reappear even after an update attempts to fix them. In this paper, we develop a taxonomy for bug types in games that is based on prior work. We 
examine 12,122 bug fixes from 723 updates
for 30 popular games on the Steam platform.  
We label the bug fixes included in these updates to identify the frequency of these different bug types, the rate at which bug types recur over multiple updates, and which bug types are treated as more severe. Additionally, we survey game developers regarding their experience with different bug types and what aspects of game development they most strongly associate with bug appearance. We find that \textit{Information} bugs appear the most frequently in updates, while \textit{Crash} bugs recur the most frequently and are often treated as more severe than other bug types. Finally, we find that challenges in testing, code quality, and bug reproduction have a close association with bug persistence. These findings should help developers identify which aspects of game development could benefit from greater attention in order to prevent bugs. Researchers can use our results in devising tools and methods to better identify and address certain bug types.
\end{abstract}


\section{Introduction}
\label{sec:intro}

Similar to any regular software, the presence of bugs in video games can cause considerable problems for game developers, including lost sales, damaged public image, and even lawsuits~\cite{Fallout76, Battlefield4}. These bugs can also impact the users--and their ability to enjoy the full range of game features--in a variety of ways. Certain bugs can cause players to lose hours of progress~\cite{PaperMario}, while others can create misunderstandings between players that might result in receiving ``abuse from angry teammates''~\cite{ApexLegends}.
Despite the negative consequences a bug can have on developers and users alike, it is common for games to release with bugs that are fixed through subsequent updates.
In 2014, for example, over a third of big-budget games ``released on Xbox One, Wii U and PS4'' received an update within 24 hours of the game's initial release~\cite{NarcisseDayOnePatch}.

The differences between the development of games and development of traditional software could explain the appearance of these bugs. Prior research has focused on identifying these differences~\cite{murphy2014cowboys, santos2018computer}. For example, games typically do not have strict functional requirements~\cite{murphy2014cowboys}. Unlike traditional software that is usually meant to accomplish a particular task, generally the primary requirement of a game is to simply be ``fun''~\cite{murphy2014cowboys}. As a result, requirements tend to be more informal and subject to change as developers continually refine the game to be more enjoyable~\cite{murphy2014cowboys, santos2018computer}. Additionally, game development teams are often more diverse in terms of team member background and expertise~\cite{murphy2014cowboys}. For example, games often contain a large number of graphical assets, which requires greater participation from designers and artists~\cite{murphy2014cowboys}.

One important difference that could explain why bugs may not get detected before releasing relates to the difficulty in comprehensively testing all aspects of a video game~\cite{murphy2014cowboys, santos2018computer}. 
Developers have difficulty writing comprehensive tests, because games can have a significantly large number of possible user interactions compared to other types of software~\cite{santos2018computer}. Players can act unpredictably when playing games, and automated testing tools struggle to replicate the full range of interactions that human players might attempt when playing a game~\cite{santos2018computer}. As a result, many games release with undiscovered bugs that only reveal themselves once customers begin playing the game~\cite{NarcisseDayOnePatch}.

One way developers address previously undetected bugs post-release is by applying updates to their games that fix these bugs~\cite{lin2017studying}.
These updates are usually accompanied by update notes, a textual description of the changes in the accompanying update~\cite{vsvelch2019resisting}.
These update notes are often published as news items on the game's website (e.g.,~\cite{brawhallaUpdate4}) or on an official online game forum (e.g.,~\cite{ESO605}).
While some updates might include content changes or additions, other updates might only attempt to fix a small number of severe bugs~\cite{lin2017studying}. Some of these smaller updates are referred to as ``hotfixes''; these updates are typically meant to remedy more pressing issues or bugs that need immediate attention~\cite{vsvelch2019resisting}.

Additionally, some bugs might reappear or recur in multiple updates. There exist cases in which developers have attempted to fix a bug in one update, only for that bug to reappear in the game anyway, necessitating another attempted fix in a later update~\cite{CoDWarzoneBug}.  
For example, in the game \textit{Warframe}, the notes for an update posted in August of 2019 included a bug fix statement that said, \textit{``Fixed inability to complete the Mastery Rank 12 test if you fall off the starting platform''}~\cite{WarframeRecurAug}. In September of 2019, the notes for a later update also included an identical line, which implies that the August update did not successfully fix the bug~\cite{WarframeRecurSep}.

Previous research has focused on creating a taxonomy for video game bugs~\cite{lewis2010went}. While this taxonomy provides a solid basis for categorizing bugs in games, a cursory examination of bug fixes described in game update notes indicated that this taxonomy did not cover all the bugs addressed in these updates.
Since the taxonomy developed by Lewis et al. was developed primarily by observing videos of game play and reading articles related to bug complaints from users~\cite{lewis2010went}, it fails to identify the bugs that are more subtle and less readily noticeable, such as bugs related to background variables that are not presented to the player.

To address the shortcomings of the existing taxonomy and to attain a deeper understanding of the types of bugs in games,
in this paper, we expand the taxonomy of bug types. We posit that improving the taxonomy will improve the chances of identifying the types of bugs missed in testing and will also improve the development pipeline in order to better identify these bugs. This will allow us to gain a deeper understanding of the association between different bug types to different game development techniques and processes.

Using this expanded taxonomy, we collect update notes for 30 popular games on the Steam platform~\cite{SteamHome, SteamStats} and categorize the bug fixes present in these updates. We analyze the frequency at which the different bug types appear in the update notes and investigate which types of bugs recur more often over multiple updates. 
Additionally, we investigate 
which types of bugs most frequently appear in urgent updates or hotfixes, as the bugs that appear in these updates are more likely to have a severe negative impact on users~\cite{vsvelch2019resisting}. Finally, we survey game developers on their experience with these different types of bugs as well as what challenges and techniques are involved in fixing these bugs.

While past research has investigated the content and timing of game updates~\cite{lin2017studying}, to the best of our knowledge, our work is the first to examine the frequency in which certain types of bugs appear and recur in game updates and the first to investigate the reasons why certain types of bugs appear more frequently than others.
Specifically, this paper addresses the following research questions:

\textbf{RQ1}: What are the most frequently fixed types of bugs through game updates?

\textbf{RQ2}: What are the most frequently recurring types of bugs?

\textbf{RQ3}: Are all types of bugs equally severe in negatively impacting the game experience?

In addition to answering these research questions, we have provided our dataset of 12,122 labelled bug fixes and a report of the survey responses for the purpose of aiding future research on this topic~\cite{replic}.

\section{Related Work}
\label{sec:relatedwork}

Prior research on video games have looked into creating a taxonomy of bugs, identifying design patterns and smells. Lewis et al. created a taxonomy for video game bugs after surveying online videos, articles and online communities dedicated to video game issues~\cite{lewis2010went}. This taxonomy serves as the basis for the taxonomy used in our study. As part of a broader investigation into the relationship between programming languages and software quality, Ray et al. developed a taxonomy of bugs based on the cause and impact of the bug~\cite{ray2014large}. 
This taxonomy was used by Pascarella et al. when classifying faults in open source games~\cite{pascarella2018video}. 
However, we found that this taxonomy was not suitable for our purposes. For one, it is not specific to games, which limits their utility for aiding game developers with the issues that are more particular to their environment. Secondly, this taxonomy is based on non-functional characteristics including Performance, Concurrency, Algorithm, Memory, Programming, and Security, and was developed by analyzing bug fix commit messages from open source projects~\cite{ray2014large}. This type of information is not available in release notes for popular non-open source games~\cite{ray2014large}.

Ampatzoglou et al. performed a case study on a collection of open-source Java games to find possible correlations between the application rate of design patterns, the defect frequency and the debugging efficiency~\cite{ampatzoglou2011empirical}. They found that certain patterns such as the Adapter pattern and Observer pattern exhibited some correlations with defects and debugging, but could not identify causal relationships~\cite{ampatzoglou2011empirical}. Borrelli et al. identified seven types of bad smells in game projects that were created with the Unity game engine and presented UnityLinter, a static analysis tool designed to help Unity developers detect these smells~\cite{borrellidetecting}.

Some researchers have focused on identifying  
differences between game development and development of other software. Murphy-Hill et al. conducted surveys and interviews with game developers~\cite{murphy2014cowboys}. Some of the findings were related to difficulties with testing; 
some game developers expressed difficulty in designing tests that could explore the state space in games, while others indicated that there was a lack of automated, low-level testing~\cite{murphy2014cowboys}. Santos et al. examined the particularities of software testing in game development, and, through the use of case studies and surveys, found a number of difficulties that game developers face while testing their software, which, in addition to challenges in test automation, included rapidly changing requirements and impracticality in testing interactive components~\cite{santos2018computer}.

There exists prior research on software release notes.
Abebe et al. investigated the types of content that appear in release notes for updates to general software. They found six types of information that are generally in release notes, including a description of the issues addressed in the release~\cite{abebe2016empirical}.

Lin et al. performed an empirical analysis of urgent updates for games on the Steam platform--including which types of games released urgent updates, the timing of these updates, and the reasons for releasing urgent updates~\cite{lin2017studying}. With respect to the reasons for releasing urgent updates, the authors created a taxonomy for these primary reasons~\cite{lin2017studying}. However, this taxonomy was not tailored specifically to bugs and included reasons that were not strictly related to bugs~\cite{lin2017studying}. When matching their taxonomy to the taxonomy of Lewis et al. some of the Lewis et al. categories fell under multiple categories proposed by Lin et al., while one of the Lin et al. categories was matched to seven of the 11 Lewis et al. categories~\cite{lewis2010went, lin2017studying}. In this paper, our taxonomy is focused on bug fixes rather than the general content of updates. Additionally, Lin et al. did not look into why certain types of bugs appear in updates, what kind of bugs recur over multiple updates and why, and how severe these different bug types are in comparison to each other. Furthermore, they did not survey game developers about their experience with game updates. These are all major components of our work. 
\section{Methodology}
\label{sec:method}

The methodology for our study included two primary components.
The first source of information encompassed update notes for popular games. The second source of data were responses from a survey distributed to game developers. 

\subsection{Data Collection}

For our dataset, we used update notes from the most popular games on the Steam platform~\cite{SteamHome}. We chose the Steam platform, because Steam is the largest vendor for games on the PC system~\cite{toy2018large}. To identify a selection of popular games, we found Steam's list of games that had the highest number of players on June 23, 2020~\cite{SteamStats}. We chose the top 30 games from this list, excluding items that were not really games (e.g.,~\cite{WallpaperEngine}). The list of games used for this paper can be found in Table \ref{tab:gameTable_CR}. This table also identifies the game's primary genre, based on the genre tags displayed on the game's Steam Store page~\cite{steamTags}.

\begin{table*}[t]
\caption{Most Popular Steam Games on June 23, 2020}
\label{tab:gameTable_CR}
\resizebox{\textwidth}{!}{%
\begin{tabular}{|l|l|c|}
\hline
\textbf{Game Title} & \textbf{Primary Genre} & \multicolumn{1}{l|}{\textbf{Number of Analyzed Updates}} \\ \hline
Counter-Strike: Global Offensive & Shooter & 36 \\ \hline
Dota 2 & MOBA & 16 \\ \hline
Destiny 2 & Shooter & 18 \\ \hline
PLAYERUNKNOWN'S BATTLEGROUNDS & Survival & 24 \\ \hline
Path of Exile & RPG & 87 \\ \hline
Grand Theft Auto V & Action & 3 \\ \hline
Tom Clancy's Rainbow Six Siege & Shooter & 16 \\ \hline
Football Manager 2020 & Simulation & 7 \\ \hline
Team Fortress 2 & Shooter & 14 \\ \hline
ARK: Survival Evolved & Survival & 80 \\ \hline
Dead by Daylight & Survival & 23 \\ \hline
Rocket League & Sports & 10 \\ \hline
Terraria & Survival & 5 \\ \hline
Rust & Survival & 14 \\ \hline
Sid Meier's Civilization VI & Strategy & 3 \\ \hline
Warframe & Shooter & 98 \\ \hline
Total War: WARHAMMER II & Strategy & 6 \\ \hline
PAYDAY 2 & Shooter & 10 \\ \hline
Europa Universalis IV & Strategy & 8 \\ \hline
The Elder Scrolls V: Skyrim Special Edition & RPG & 1 \\ \hline
The Elder Scrolls Online & RPG & 25 \\ \hline
Hearts of Iron IV & Strategy & 7 \\ \hline
SMITE & MOBA & 12 \\ \hline
Unturned & Survival & 19 \\ \hline
Sid Meier's Civilization V & Strategy & 5 \\ \hline
Black Desert Online & RPG & 49 \\ \hline
Brawlhalla & Fighting & 15 \\ \hline
War Thunder & Simulation & 92 \\ \hline
Euro Truck Simulator 2 & Simulation & 3 \\ \hline
FINAL FANTASY XIV Online & RPG & 17 \\ \hline
\end{tabular}
}
\end{table*}

Once we had our collection of games, we manually extracted the update notes. For each game, we collected all update notes in the one-year period leading up to the date of the game's most recent update. 
For example, the most recent update we collected for the game \textit{Counter Strike: Global Offensive} was released on June 22, 2020, which meant the earliest update we collected for this game was released on July 16, 2019.
Another game on the list, \textit{Sid Meier's Civilization V}, was much older, with its most recent update being released on October 30, 2013, which meant the earliest update we collected was on March 13, 2013.

Each game on Steam had its own channel that displayed news and updates about the game, including some update notes. We found this source was incomplete for many games, however. Not all of the game's updates were always posted to the Steam channel. For instance, smaller updates that only contained one or two lines--such as hotfix updates--seemed less likely to appear on the Steam channel. For most games, we used the updates notes from the official game or developer web site (e.g.,~\cite{brawhallaUpdate4}) or online forums (e.g.,~\cite{ESO605}), relying on the Steam update channel only if there was on other option.

Once we collected the update notes, we used a script to collect the text from the updates related to bug fixes. The text of each update note was split into smaller segments of text based on the line breaks of the source page formatting (referred to as ``lines'' going forward). Under this scheme, each paragraph or list item was treated as an individual line.  
A formative analysis of the update notes  
revealed certain words or phrases that  
were frequently present in bug fix lines, such as ``fix'' and ``corrected''. We applied a simple filtering scheme that scanned through all our update notes and flagged lines with these terms.
Each of these flagged lines was recorded in our data alongside supplemental information that included the game the update containing the line came from, the date of the update, and the ID of the update (for cases in which multiple updates were released in a single day). The number of analyzed updates for each game can be seen in Table \ref{tab:gameTable_CR}. An update was considered analyzed if it contained at least one labelled bug fix, as described in the following section.

\subsection{Types of Bugs}

The next step was to label the bug fixes in our data. Initially, our goal was to label the bug fixes based on the 11 bug types identified by Lewis et al.~\cite{lewis2010went}. However, we identified that there were bug-fixes that did not fit in any one of the existing types. Hence,  
two of the authors
jointly applied open coding~\cite{seaman1999qualitative}.
First, we generated short descriptions of the initial 11 bug types based on the descriptions from Lewis et al.~\cite{lewis2010went}. As we labelled bug fixes from our data, if we found a bug fix that did not fall under a currently available label, we created a new label.
Additionally, we would search through previously labelled bug fixes to check if any of these fixes better fit under the new label. This step additionally helped remedy instances in which a bug could conceivably fall under multiple categories, a problem Lewis et al. also encountered~\cite{lewis2010went}. In cases like these, we chose which category best described the core impact of the bug. For example, one bug fix for the game \textit{Black Desert Online} said, \textit{``fixed the issue where running left or right was impossible when transformed into an immortal persimmon knight''}. While part of this bug related to incorrect position of a game object, the inability to perform an action seemed to be the core problem.
Therefore, the \textit{Action} label was applied instead of the \textit{Position of Object} label.

Finally, after labelling the full collection of bug fixes, we reviewed and consolidated our categories, combining categories that could be classified as sub-types of another category.
During this step, we also discarded lines that did not actually contain bug fixes and lines in which the nature of the bug was unclear and could not be categorized. During the coding process, there were no disagreements between the authors.
These bug types
are described in Table \ref{tab:bugTypes2}. 
The nine new bug types are marked in the table with an asterisk (*).

In total, 12,122 lines were labelled under this taxonomy; these labelled lines came from 723 different updates. The labelled dataset of these lines is available online~\cite{replic}.

\begin{table*}[t]
\centering
\caption{Types of bugs found in update notes}
\begin{tabular}{|p{2.5cm}|p{5.5cm}|p{2.5cm}|p{5.5cm}|}
\hline
\textbf{Category} &
  \textbf{Description} &
  \textbf{Category} &
  \textbf{Description} \\ \hline
Action &
  Errors in the ability/inability to perform actions.
  &
  Game Graphics &
  A certain visual aspect of the game world is being rendered incorrectly.
  \\ \hline
Artificial Intelligence &
  An NPC or AI does not behave in the intended manner. &
  Implementation Response &
  There are errors regarding the game's interactions with hardware or base software that ends up affecting performance.
  \\ \hline
Audio* &
  Game audio is not playing correctly.
  &
  Information &
  Information about the game world is not conveyed to the player correctly.
  \\ \hline
Bounds &
  An object exists outside the intended game space, such as being outside the game map or in a location that should not be accessible. &
  Interaction Between Object Properties* &
  Two or more object properties do not behave properly when interacting with each other.
  \\ \hline
Camera* &
  There is an issue with game camera through which the player views the game world. &
  Interrupted Event &
  An action in game that was previously operating is terminated abruptly against expectations, or the action was not interrupted when it should have been. \\ \hline
Collision of Objects* &
  Objects do not behave properly when they collide or make contact with each other. &
  Object Persistence* &
  Game objects do not correctly enter or exit the game world.
  \\ \hline
Context State &
  Objects do not enter and exit states properly. &
  Position of Object &
  An object is not in the correct position or orientation within the game space over time.
  \\ \hline
Crash* &
  The game crashes or is otherwise forced to close. &
  Triggered Event* &
  Improper triggering of game events (Event A causes Event B).
  \\ \hline
Event Occurrence &
  Discrete events do not occur at the correct rate or order.
  &
  User Interface* &
  The appearance/position/behavior of UI elements are incorrect. \\ \hline
Exploit* &
  Players are able to improperly and unfairly gain game benefits through behavior unintended by developers. &
  Value &
  An in-game variable is not set to the correct value. \\ \hline
\end{tabular}
\label{tab:bugTypes2}
\end{table*}

\subsection{Bug Recurrence}

The next part of our process required us to identify all recurring bug fixes that appeared in multiple updates for a single game. Due to the large number of updates and the large number of bug fixes, manually searching through the data to find all recurring bugs would be prohibitively time consuming. At the same time, we discovered a purely automated approach could also lead to problems. A cursory analysis of repeating bug fixes in the data revealed that some repeating lines were generally written; it was unclear whether or not the fix was applying to the same bug each time. 
For example, the game \textit{ARK: Survival Evolved} had 5 updates that included the line ``fixed a server crash''. In this case, it was not apparent whether this line was referring to the same bug each time or different bugs that each caused crashes. Conversely, two updates for the game \textit{Dead by Daylight} both contain the same line: ``\textit{fixed an issue that caused a crash when changing the graphic quality settings during a match in the red forest maps}''. Here, the apparent cause and effect of the bug are identical; changing certain settings in a certain context causes the game to crash. In this case, both lines are likely referring to the same bug. 
We were skeptical that an automated tool would be able to differentiate between these two types of recurring bug fixes, so we chose not to use a purely automated approach either.

We ultimately decided on a hybrid approach. 
We would use an automated approach to limit our analysis to a smaller collection of bug fixes and then manually evaluate these lines to find the true recurring bugs. For 
this first step,
we performed cosine similarity analysis~\cite{salton1988term} 
between bug fix lines. For each bug fix line from a game's updates, we compared that line to each line in all subsequent updates for the game. 
If two lines had a similarity score of at least 90\%, they were treated as a potential match and grouped together for manual review.
Our small-scale formative analysis showed that a similarity threshold of 90\% was sufficient to minimize false positives. A line could have multiple matches if it appeared in more than two updates; all instances of these bugs were grouped together. Once we collected all potential matches, we manually inspected them and removed all false positives as well as those matches in which it was unclear whether the matching lines were really referring to the same bug.
After filtering the potential matches, we were able to collect information on the types of bugs that recurred over multiple updates.

\subsection{Bug Severity}

To identify which bugs were treated as severe, we used metrics employed by Lin et al. to identify urgent updates~\cite{lin2017studying}. Under these metrics, updates were considered urgent if they were: identified as hotfixes within the notes for the update, 0-day updates (same day as the last update), or deployed sooner than normal, based on the game's regular update schedule~\cite{lin2017studying}.

In addition to these metrics, we also looked at the length of the update notes. 
Based on our own observations of the data, it appeared as if updates addressing urgent bugs typically had shorter update notes than regularly scheduled updates. This is likely because urgent updates are only intended to fix a small number of critical issues that cannot wait until the next major update.
To find the urgent updates in our data, 
we used a script to parse through each update in our data to check the following information: whether the update notes contained language referring to the update as a hotfix, the number of days between the current update and the last update, and the number of non-whitespace lines in the update notes.

For the self-identified hotfixes, we searched for updates that contained the word ``hotfix'' in the first twenty lines of the update text. This was to help avoid false positives, because updates that self-identified as hotfixes would use the language near the beginning of the update text. When the word appeared further into the text, this was typically because the update was referring to another hotfix update, such as a past update or a planned future update. After running our script, we then manually reviewed the resulting updates and filtered out any remaining false positive updates and marked the remaining updates as urgent. For the 0-day updates, we marked as urgent all updates that had a value of 0 days since the last update.

For the last two metrics (number of days and number of non-whitespace lines), we used Double MAD (Median Absolute Deviation)~\cite{lin2017studying, doubleMAD} to find the outliers below the median for both the days since the last update and for the update size. Updates that fit either of these criteria were also marked as urgent. We used the same threshold value of 2 that was used by Lin et al. in their paper~\cite{lin2017studying}.

Once we identified all the urgent updates in our data, we flagged all the  bug fixes lines from our data that appeared in an urgent update. An urgent update is generally intended to fix ``problems that are deemed critical enough to not be left unfixed until a regular-cycle update''~\cite{lin2017studying}.
As such, the bug fixes included in these urgent updates are likely to have severe negative impacts on the game experience that require more immediate attention. For our purposes then, if a bug fix was included in an urgent update, we treated it as a severe bug. We used the findings from this process to calculate the severity rate of each bug type, as explained in Part \textit{C} of Section IV.

\subsection{Survey}

\textit{Protocol:} We created a 12-minute survey using the Qualtrics survey system~\cite{qualtricsHome} 
to assess game developer perceptions of the types of bugs that appear in game update notes. First, the survey asked participants certain demographic questions, such as the years of game development experience and the number of people in their organization. We also asked participants if they had experience fixing bugs for updates to video games. The next section asked participants about their experience with the different bug types in our taxonomy. For each of the twenty bug types, we asked participants to rate on a 5-point Likert scale their level of agreement that the bug type was likely to have a severe negative impact on the game experience. 
Participants were then informed of the three most frequent bug types across all updates and were asked if these results matched with their own experience. They were also offered to explain why they believed these bug types (and other bug types they commonly encounter) appear so frequently in game updates. Similar questions were also asked for the frequently recurring bug types. Finally, participants were asked about the aspects of game development related to bug recurrence and the challenges in identifying and fixing bugs.

\textit{Respondents:} To find participants, 
we identified GitHub~\cite{GitHub} repositories for games and game-related tools that had been marked as noteworthy by other GitHub users. We sent the survey to GitHub users who had contributed to these repositories and who had made their email addresses visible to other users. Additionally, we sent the survey to developers of the games examined for this study whose email addresses were publicly listed on the developers' web sites. We also made the link to the survey available to a collection of game developers with the intention that they would apply snowballing to spread the survey to other developers~\cite{kelley2003good}.

The survey was sent to 630 contacts. 
16 of these invitations were not delivered successfully. Three invitations generated automatic replies suggesting that the recipient was unable to receive the survey, and one recipient indicated that he did not possess a game development background. Overall, 610 survey invites were delivered, and we received 47 applicable responses. This gave us a response rate of 7.70\%. Other important studies in the software engineering field have reported response rates ranging between 5.7\%~\cite{passos2018study} and 7.9\%~\cite{FlavioMedeiros_2018}, indicating that our response rate for this study is satisfactory. The report of the survey results is available online~\cite{replic}.

\section{Results}
\label{sec:result}
\subsection{Frequent Bug Types in All Updates}

\begin{figure}[ht]
    \includegraphics[width=\columnwidth]{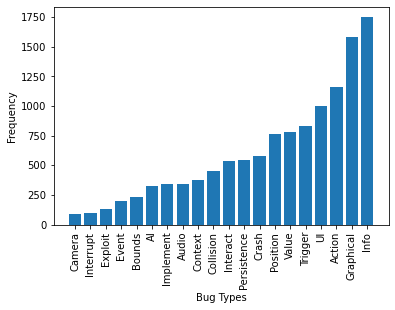}
    \caption{Frequency of Bug Types Across All Updates}
    \label{fig:freqGen}
\end{figure}

Figure \ref{fig:freqGen} shows the frequencies at which each of the 20 bug types appeared as a fix in any game update from our data. \observation{The most frequently occurring bug types were \textit{Information}, \textit{Game Graphics}, and \textit{Action}.} The three least frequent bug types were \textit{Camera}, \textit{Interrupted Event}, and \textit{Exploit}. All three of the most frequent bug types were bug types identified by Lewis et al.~\cite{lewis2010went}.
Meanwhile, for the three least frequent bug types, only \textit{Interrupted Event} was one of the bug types identified by Lewis et al.~\cite{lewis2010went}. 

The two most frequent bug types, \textit{Information} and \textit{Game Graphics} cover largely non-interactive elements of the game. This runs somewhat contrary to findings found by Santos et al. that suggest some of the difficulties in testing games are due to challenges in testing interactive elements~\cite{santos2018computer}. If the most frequent bug types are not necessarily interactive in nature, perhaps automated testing might be more effective at catching these more common types of bugs. 

For each of the 
three most frequent bug types, survey 
respondents were asked if they frequently fixed bugs of that type in game updates. 57\% of respondents said they frequently fixed \textit{Information} bugs, 40\% said they frequently fixed \textit{Game Graphics} bugs, and 67\% said they frequently fixed \textit{Action} bugs. 

When asked why certain bug types might appear frequently, respondents gave a variety of responses. 
For \textit{Information} bugs, some respondents 
referenced priority concerns, such as the one in the following survey response.
\begin{itemize}

\item \textit{``Developing mechanics is the main focus and usually information related components are pushed down to the last of the pipeline and as deadline approaches, they are not done right.''}
\end{itemize}

Other respondents noted that the correct information may be difficult to identify if that information is the result of various calculations involving \textit{``complex states and interactions''}. 
This information could also be \textit{``buried in data files''} that testers may not be able  
to locate. Incorrect information might be missed, because the correct information may not be 
accessible.

\observation{Survey explanations for the frequency of  \textit{Information} bugs include: low priority compared to other aspects of the game and difficulties in finding the correct information.}

With respect to \textit{Game Graphics} bugs, the most cited reason by respondents to their frequency in updates was due to the complexity of the subject.
\begin{itemize}
\item \textit{``Computer graphics is one of the most complex disciplines to work with and need highly experienced programmers to make sure that everything just work.''}
\end{itemize}

Respondents also cited the large number of cases that need to be tested. 
\begin{itemize}
\item \textit{``It's fairly simple to miss visual corner cases in testing.''}
\end{itemize}

Additionally, some responses seemed to speak to the difficulties in properly assessing the quality of game graphics. One respondent pointed out that game graphics can be subjective. What looks like a graphical bug to one tester may not look like a bug to another tester. Another respondent mentioned that the correct graphical qualities of the game might be highly specific and require fine-tuning to get them right.
\begin{itemize}
\item \textit{``Graphics can be delicate. It can take a lot of tweaking to get something to look just right''}
\end{itemize}
One response noted that  
graphics 
are considered a lower priority to fix before production than other game elements.

\observation{Survey explanations for the frequency of  \textit{Game Graphics} bugs include: complexity, large size of testing state space and evaluation of graphics quality.}

For \textit{Action} bugs, many of the respondents spoke to complexity in testing as a reason these bugs might appear frequently in updates. Each possible action gives the player a new way to interact with the game world. As the number of possible interactions increases, there is a greater chance that these interactions might conflict with each other. Testing all possible conflicts can be difficult to do within the development timeline, especially when dealing with actions that may or may not be executable depending on the game state. 
\begin{itemize}
\item \textit{``Games often have a large number of available actions, only some of which are available at a given time; the results of actions often depend on context (and sometimes randomness); and actions may depend on or be changed by previously-performed actions.  While testing every possible action might be feasible, testing every possible sequence of actions in every possible sequence of contexts is usually combinatorially impossible.''}

\end{itemize}

\observation{Survey explanations for the frequency of \textit{Action} bugs include: increasing complexity with additional possible actions and large size of testing state space.}

Respondents were also asked if they frequently fixed bugs of any other type from our taxonomy and why these bugs might be frequent. \textit{Crash} bugs were identified as frequent by two respondents, with one respondent attributing this frequency to \textit{``not properly checking for input sanity''}. Another respondent mentioned \textit{Exploit} bugs, stating \textit{``players are good at finding them''}. Another respondent who worked with multiplayer games referred to bugs related to desynchronization of players. Under our taxonomy, this bug would fall under the \textit{Implementation Response} label. Another user mentioned \textit{Bounds} bugs as a frequent concern, especially for 3D games, explaining that the collision and physics systems in a game tend to produce edge cases somewhere that leads to bounds issues.

\subsection{Recurring Bugs}

\begin{figure}[h]
    \includegraphics[width=\columnwidth]{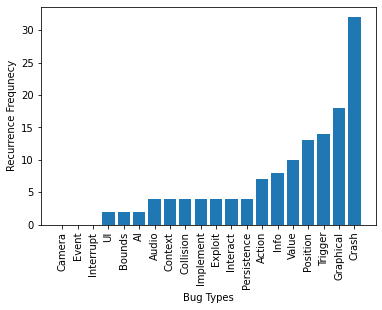}
    \caption{Frequency of Recurring Bug Types}
    \label{fig:recur}
\end{figure}

Figure \ref{fig:recur} shows the frequency in which bugs of a certain type would recur over multiple updates. \observation{The bug types that recurred the most frequently over multiple updates were \textit{Crash}, \textit{Game Graphics}, and \textit{Triggered Event}.}
The bug types that recurred over multiple updates the least frequently were \textit{Camera}, \textit{Event Occurrence}, and \textit{Interrupted Event}; none of these bug types had any instances of bugs that recurred over multiple updates.
One of the most frequently recurring bug types--\textit{Game Graphics}--and two of the least frequently recurring bug types--\textit{Interrupted Event} and \textit{Event Occurrence}--came from the taxonomy of Lewis et al.~\cite{lewis2010went}.

Regarding 
the top three frequently recurring bug types, survey respondents were asked if they frequently fixed recurring bugs that fell into each of these categories. 48\% of respondents said they frequently fixed recurring \textit{Crash} bugs, 21\% said they frequently fixed recurring \textit{Game Graphics} bugs, and 24\% said they frequently fixed recurring \textit{Triggered Event} bugs. 

For each of these top three bug types, survey respondents were offered to explain why these bug types recur over multiple updates more frequently. 
When it came to \textit{Crash} bugs, multiple respondents cited the difficulties in identifying the actual root cause of the crash.  
\begin{itemize}

\item \textit{``When fixing a crash bug, it is easier to find the single caller that passed bad data in the crash report than it is to ensure that all callers pass good data.''}
\end{itemize}

Other respondents spoke to how
other changes to the game can easily cause the crash to reappear even after fixing it.
\begin{itemize}
\item \textit{``Due to how interconnected most things in games are, it is easy to reintroduce a crash after initially fixing it''}
\end{itemize}

\observation{Survey explanations for the frequent recurrence of  \textit{Crash} bugs include: reproducing bugs and reappearance after other changes.}

With respect to why \textit{Game Graphics} bugs would recur frequently, the responses were more mixed. One respondent mentioned that game graphics can be \textit{``inter-dependent and cascade through fixes''}, while another expressed experience with having to deal with unforeseen edge cases.

A different respondent reiterated the delicate nature of game graphics. A previous fix to a graphical component of the game might not have been sufficient, and further tweaking might be required to get the graphics just right. Another user pointed out that \textit{``Graphics bugs are the most apparent thing for the player to notice''}. Building off these responses, it is possible that--since the quality of game graphics can be subjective, and because players more readily notice the quality of game graphics--these kinds of bugs recur more frequently because players are more likely to notice graphics that do not meet their personal quality standards, which could mean they are more likely to express this dissatisfaction to developers.

\observation{Survey explanations for the frequent recurrence of  \textit{Game Graphics} bugs include: inter-dependence of graphics, unforeseen edge cases, and the specific nature of graphical components.}

Because few respondents claimed to see frequently recurring \textit{Triggered Event} bugs, there were not many explanations as to why these bugs might recur frequently over multiple updates. Still, some responses speak to the large size of the test space.
\begin{itemize}
    \item \textit{``Events can often be triggered in many different ways; making sure every way works is tricky and time-consuming.''}
\end{itemize}

\observation{Survey explanations for the frequent recurrence of  \textit{Triggered Event} bugs include: large size of testing state space.}

When asked if respondents had seen any other 
frequently recurring bug types,
one respondent pointed to \textit{Action} bugs, stating that these bugs recur often for the same reasons as \textit{Crash} bugs. 
Other respondents spoke more broadly, identifying reasons why 
bugs in general might recur over  
updates. 
\begin{itemize}
\item \textit{``Sometimes bugs recur because some careless guy removes your fix when merging his/her code.''}
\item \textit{``It has to do with whether management allows us the time for testing before they push out the newest code.''}
\end{itemize}

\subsection{Bug Severity}

\begin{figure}[h]
    \includegraphics[width=\columnwidth]{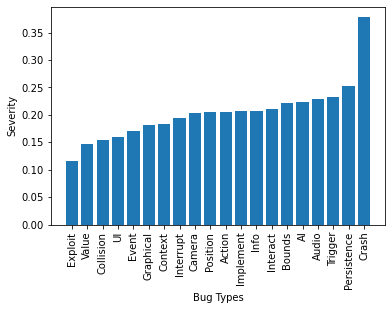}
    \caption{Severity of Bug Types}
    \label{fig:severity}
\end{figure}

We calculated the severity of each bug type by finding the proportion of bug fixes from the data that appeared in an update marked as urgent. For each bug type, we divided the frequency of the bug type in an urgent update by the frequency of the bug type in all updates, as expressed below:

\begin{align}
    Bug Type Severity &= \frac{Bug Type Freq(Urgent Updates)}{Bug Type Freq(All Updates)}
\end{align}

Figure \ref{fig:severity} shows the resulting severity rates of each bug type. \observation{Based on the update data, the bug type with the highest severity was \textit{Crash}. The next most severe bug types were \textit{Object Persistence} and \textit{Triggered Event}.} \textit{Crash} bugs appeared 580 times across all updates and 219 times in urgent updates. 38\% of the occurrences  
were in urgent updates, the greatest proportion for any bug type.

\begin{figure}[ht]
    \includegraphics[width=\columnwidth]{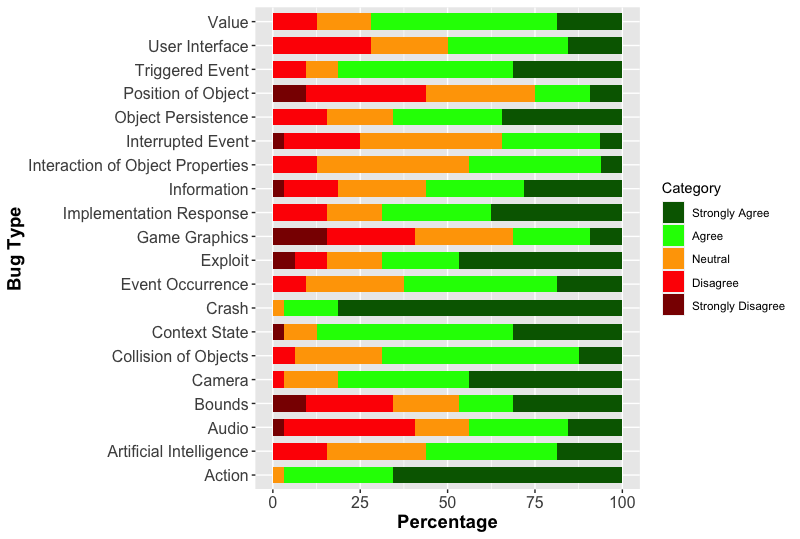}
    \caption{Agreement Level to Whether Selected Bug Types are Likely to Have a Severe Negative Impact on Game Experience}
    \label{fig:sevSurvey}
\end{figure}

Figure \ref{fig:sevSurvey} depicts the perceptions of bug severity among survey participants. For each bug type \textit{X}, respondents were asked to state the level of agreement with the statement ``\textit{X} bugs are likely to have a severe negative impact on the game experience.'' \observation{The bug types that generated the largest number of \textit{Strongly Agree} responses were \textit{Crash}, \textit{Action}, and \textit{Exploit}.}

With respect to \textit{Crash} bugs, the results from the update data and from the survey appear to match; \textit{Crash} bugs appear to be the most severe bugs in both data sources. 
\textit{Action} 
bugs, the second-most severe type from the survey, do not rank as highly when looking at the mined data seen in Figure \ref{fig:severity}, however. Interestingly, \textit{Exploit} was the bug type that received the third-most \textit{Strongly Agree} responses, yet it had the lowest severity rate from the mined update data. The reason for this discrepancy is not readily clear. It could be due to a difference between the perceived impacts of these bugs versus the actual impact. This may be a worthwhile avenue for future research.

\subsection{Bug Fix Association with Game Development Processes}

Survey respondents were asked \textit{``In your opinion, what aspects of game development (i.e. testing, designing, etc.) are associated with bug recurrence in your current project or in other game projects in which you participated?''}  
16 respondents mentioned testing as a component of game development that had a strong association with bug recurrence. Some of these responses mentioned how a lack of testing in general was linked to bug recurrence, while other responses spoke about the lack of certain kinds of testing, such as automated testing, integration testing, and cross-platform testing.
\begin{itemize}
    \item \textit{``Lack of automated testing definitely increases bugs in general''}
    \item \textit{``The testing only checks the current feature/bugfix and not how it interacts with everything else''}
\end{itemize}

Four respondents claimed that game design had a strong link to bug recurrence.

\begin{itemize}
    \item \textit{``Good design helps avoid bug recurrence in the first place. Testing can be used as a fail-safe''}
\end{itemize}

Four respondents pointed to code quality/coding as an important component connected with bug recurrence.
\begin{itemize}
    \item \textit{``Code quality is often a huge factor for bugs.  Badly designed code interacts badly with other pieces of code, and conflicts can result easily, causing bugs.''}
\end{itemize}

Other aspects of game development provided in the responses included the game architecture, quality assurance, a lack of consistency checks, and the addition of new features.
\begin{itemize}
    \item \textit{``In my experience, new gameplay features often lead to bug recurrence, when a previous bug was "solved" by addressing the particular feature interaction that caused the bug rather than making those features more robust in general.''}
\end{itemize}

\observation{The aspects of game development most frequently linked to bug recurrence were: testing, game design, and code quality.}

Finally, respondents were asked what they believed were the main challenges in identifying and fixing bugs in video game development. Like the previous question, 22 respondents provided answers.
Eight responses pointed to challenges with conducting adequate testing.
\begin{itemize}
    \item \textit{``too many possibilities of interaction/combination to test them all thoroughly''}
\end{itemize}

Seven responses mentioned difficulties in identifying the origin of bugs or reproducing bugs.
\begin{itemize}
    \item \textit{``Reproducing a bug often requires a large time investment to reach the state in which it is apparent. Games specifically are huge loosely connected systems; bugs often depend on more than one of these systems having a certain state.''}
\end{itemize}

Six responses focused on code quality. There was a particular emphasis on the importance of well-written code that does not cause conflicts elsewhere in the game.
\begin{itemize}
    \item \textit{``The hardest part is finding an appropriate solution to a bug that isn't hacky and is future-proof to later additions to the code that might break the bugfix again.''}
\end{itemize}

Two responses 
mentioned 
development schedules.
\begin{itemize}
    \item \textit{``Lack of QA, faulty development schedule, low focus on code quality.''}
\end{itemize}

Two responses mentioned unforeseen bugs that only appear in particular situations, such as bugs arising out of specific hardware setups.
\begin{itemize}
    \item \textit{``Bugs that occur only with some OSes or hardware''}
\end{itemize}

Two responses explained that the choice of programming language can lead to issues.
\begin{itemize}
    \item \textit{``Using languages with a weak memory model such as C++ makes it hard to identify bugs caused by memory corruption errors such as use-after-free.''}
\end{itemize}

\observation{The most frequently mentioned main challenges to identifying and fixing bugs in video games were: inadequate testing, reproducing bugs, and code quality.}
\section{Discussion}
\label{sec:discussion}

\subsection{For Researchers}

\textbf{Prioritizing of Bug Types:} Our results identify which types of bugs appear in updates more frequently than others, which bugs recur over multiple updates more frequently, and which bug types are more severe in negatively impacting the game experience for players. When comparing frequency to severity, these results could help illuminate which types of bugs researchers might prioritize when devising techniques to prevent or fix certain types of bugs. For example, though \textit{Information} bugs were the most frequently occurring bug type across all the update notes, this bug type was not as highly ranked with respect to recurrence and severity, as seen in Figures \ref{fig:severity} and \ref{fig:sevSurvey}. The results seem to indicate that even if \textit{Information} bugs are common in game updates, they do not seem to recur as often as other bugs, and they are not seen as having as much of a severe negative impact on the game experience.  
Our results imply that \textit{Crash} bugs could probably use a greater focus from researchers. Though these bugs were not the most frequent bug type in general, this was the bug type that had the highest recurrence rate and the highest severity rate in our data. Additionally, the survey results also indicate that this bug has a high level of severity. Using the findings from this paper, researchers can develop specialized tools and methods to specifically target the more problematic bug types.

Our survey results also provide some explanations as to why some of these certain bug types stand out more than others. From our results, we have found that some explanations appear multiple times and in multiple contexts for different bug types. The large size of the testing state space, for example, is provided as an explanation for why both \textit{Game Graphics} bugs and \textit{Action} bugs are fixed frequently in game updates and for why \textit{Triggered Event} bugs recur frequently over multiple updates. A deeper investigation into these common causes of bug issues and approaches into remedying them might be a worthwhile subject for future research.

\textbf{Game Development Processes:} The findings regarding bug frequency, recurrence, and severity also have implications  
relating to the game development process. Testing, code quality, game design, and bug reproduction were all aspects of game development that seemed to have a strong association with bugs, according to the survey results. While some research exists on the testing process in game development~\cite{santos2018computer}, there is room to dig even deeper. Multiple respondents cited how issues could arise when different components of the game interact with each other in an unintended manner. Researchers could focus on this issue of testing the interactions between game components in future research. Additionally, there does not appear to be much research into the other three processes within the game development context. Future research can focus on devising approaches and methods focused on improving these aspects of the game development process so that they are less likely to lead to the appearance of bugs.

For example, reproducing bugs was one of the most widely cited challenges in identifying and fixing bugs in video games. 
Reproduction is difficult, because bugs might be caused by a specific interaction of game states and properties that may not be evident to whomever experiences the bug. The lack of information about the bug's true cause leads to challenges in 
fixing the bug and a greater chance that the bug will reappear in a future release.  
Indeed, this challenge with reproducing bugs was one of the major reasons from respondents for why \textit{Crash} bugs recur frequently over 
updates. To combat this problem, researchers could focus on devising tools to better collect information about the game state so that developers are better equipped to find the true cause of these bugs.

\subsection{For Practitioners}

\textbf{Prioritizing of Bug Types:} The results can help game developers identify which types of bugs may need more attention than others. As described above, developers can evaluate the frequency, recurrence, and severity information from this paper to decide which bugs may need more attention when testing or fixing bugs.

\textbf{Game Development Processes:} Similar to researchers, game developers can take advantage of the results relating to aspects of the development process that have a 
link to the bug occurrence.
For instance, the results 
suggest that difficulties in the testing process are linked with the appearance of bugs. While the size of the game state space can make comprehensive testing difficult, the responses also seem to indicate that 
testing 
may also be too narrow in scope. Many respondents mentioned issues where testing does not involve broader interactions with other components of the system. Furthermore, the interactions between system components was was given as a reason for the recurrence of \textit{Crash} bugs. These results suggest that developers might benefit from a greater focus on multi-component interaction testing. The game space might be too large to test everything, but developers can focus on creating more tests targeting high-priority bug types. Additionally, they could emphasize using tests that incorporate multiple interacting components of the system, as opposed to testing components in isolation.

Code quality was also provided both as an aspect of game design associated with bug recurrence and as a challenge in identifying and fixing bugs in video games. Poor code quality increases complexity of the program, and it also can lead to conflicts with other components of the system, ultimately resulting in bugs. The interaction of code modules in games seems to be an important component of game development that is related to both the testing process and the coding process. Implementing practices to enforce higher code quality could end up benefiting both of these processes.

Our results point to different possible parts of the game development process 
associated with bugs. By implementing practices to improve these processes, developers might be able to mitigate the rate at which bugs appear or recur.
\section{Threats to Validity}
\label{sec:threats}

With the intent to make sure our results could be generalized, we tried to collect a large amount of update notes from a high number of games. We collected the top 30 games with the highest player counts on the Steam platform in order to collect a sample of games that represents the range of video games currently available. It is possible however, that the games studied and the update notes analyzed are not representative to the entire body of game update notes.

It is possible that survey respondents may not have understood the different bug types when answering questions regarding their experiences with these bugs. To avoid this, we provided an explanation of each bug type in the survey that described the bug and included an example bug fix from the update notes that belonged to that category.

There was a risk that the bug types in our taxonomy might be influenced by bias or that they might not account for all types of bugs. To reduce this bias, we created a taxonomy that was based on prior work 
\cite{lewis2010went}.
Additionally, the large number of bug fixes (12,122) we analyzed should help ensure that our final taxonomy is complete.

In identifying severe bug fixes in the update notes, there is a possibility that our mechanisms to identify urgent updates might not have captured the correct number of relevant updates. Our methods could have been under-inclusive or over-inclusive. To avoid this, we based our approaches off of approaches used in prior research~\cite{lin2017studying}.

Additionally, our script to identify the bug fixes in our update notes might not have identified all bug fixes. From our own manual analysis, we found common repeating words and phrases that were used in the lines containing bug fixes. We accounted for these terms when creating our script. In order to test the script's efficiency, we inspected the code of our script, ran tests with the script, and tested the script on update notes with known results. It is possible, however, that some bug fix lines may not have included these terms.


\section{Conclusions and Future Work}
\label{sec:conclusion}

In this paper, we collected the update notes from popular games on the Steam platform. We created a taxonomy of bug types  
that we used to label the bug fixes included 
in the update notes. We found that the bug types that appeared the most frequently in update notes were \textit{Information}, \textit{Game Graphics}, and \textit{Action}. The bug types that recurred the most often over multiple updates were \textit{Crash}, \textit{Game Graphics}, and \textit{Triggered Event}, and the bug types with the highest rate of severity were \textit{Crash}, \textit{Object Persistence}, and \textit{Triggered Event}. We surveyed game developers who corroborated that \textit{Crash} bugs were likely to have a severe negative impact on the game experience. The survey responses also  
suggested that challenges in testing, bug reproduction, and code quality had a strong association with the occurrence of bugs in game updates. In particular, problems arising from the interaction between different components was a widely cited issue.

These results can help game developers identify which types of bugs to pay more attention to when testing and fixing bugs. Developers can also use these results to help adjust practices related to game development process in order to better prevent, identify, and fix bugs. In addition, we have made our dataset of labelled bug fixes and survey responses available online to aid in future research on this topic~\cite{replic}.

Researchers can take advantage of these results in order to develop tools or methods that can target specific bug types that are more likely to severely impact the game. There is room for future work that can identify aspects of game development that might benefit from specialized tools or methods that address some of the challenges provided in the survey. For example, multi-component interaction testing is one area that could benefit from future research. Our study takes the first step towards fulfilling these goals.

\bibliographystyle{IEEEtranS}


\end{document}